*Article*

# Defect-Width-Tunable Resonant Elastic-Wave Transmission in Micro-Pillar Arrays

**Alok Pokharel[1,2]***

1 Independent Researcher, Sydney, NSW, Australia
2 ORCID: 0000-0002-3975-1499
* Correspondence: alokpokharel17@gmail.com

**Abstract**

Periodic micro-pillar lattices integrated on elastic substrates provide a promising platform for controlling elastic-wave propagation through localized resonant interactions. In this work, defect-engineered resonant transmission in periodic tungsten micro-pillar arrays deposited on silicon substrates is numerically investigated using finite-element simulations. Two-dimensional frequency-domain analyses were performed to evaluate the influence of one-, two- and three-pillar defects on elastic wave transmission characteristics. The results reveal strong defect-width-dependent transmission modulation together with the localized resonant elastic-wave redistribution within the periodic lattice. Full three-dimensional simulations further confirm the presence of defect-sensitive resonant localization and modified elastic-wave transport pathways. Floquet dispersion analysis of a periodic unit cell reveals multiple nearly flat resonant branches associated with low-group-velocity elastic-wave modes, indicating predominantly subwavelength locally resonant behavior. A comparative study between tungsten and copper resonators demonstrates enhanced resonant confinement in tungsten-based structures due to their larger inertial contrast with the supporting substrate. The proposed defect-engineered micro-pillar lattices provide an effective approach for frequency-selective elastic-wave control and localized resonant wave manipulation in elastic metamaterial systems.

**Keywords:** Phononic crystals; Elastic metamaterials; Local resonance; Micro-pillars; Defect engineering; Floquet dispersion; Elastic-wave transmission

## 1. Introduction

Phononic crystals and elastic metamaterials have attracted considerable attention for their ability to manipulate elastic and acoustic wave propagation through periodic structural design [1]. Such systems can exhibit frequency-selective transmission, wave localization, slow-wave propagation, and bandgap formation resulting from Bragg scattering or locally resonant interactions. In particular, locally resonant elastic metamaterials composed of resonant inclusions integrated on elastic substrates have demonstrated strong wave attenuation and flat-band behavior at wavelengths significantly larger than the lattice periodicity [2]. These properties have motivated extensive research toward applications in vibration isolation, acoustic filtering, waveguiding, and on-chip elastic-wave control.

Building upon these fundamental acoustic properties, increasing attention has been directed toward structural configurations capable of enhancing elastic-wave control through localized resonant interactions. Among these approaches, pillared elastic plates consisting of periodic resonator arrays integrated on thin substrates have emerged as promising platforms for manipulating guided elastic waves [1–3]. The interaction between localized resonator modes and elastic waves propagating through the supporting substrate can significantly modify the dispersion characteristics of the structure, leading to flat-band behaviour, low-group-velocity propagation, and frequency-selective transmission. Furthermore, the intentional introduction of structural defects, such as missing or modified resonators, breaks the periodic symmetry of the lattice and enables localized resonant elastic-wave behaviour [4]. Such defect-engineered resonant structures provide an effective framework for investigating frequency-dependent transmission modulation and localized elastic-wave transport in subwavelength periodic systems.

In this work, defect-width-dependent resonant elastic-wave transmission in periodic micro-pillar arrays integrated on a silicon substrate is numerically investigated using finite-element simulations. Two-dimensional side-view models are first employed to examine defect-induced transmission modulation and localized resonant behavior in finite lattices containing missing-pillar defects. Full three-dimensional simulations are subsequently performed to validate the persistence of the observed transmission characteristics in realistic finite micro-pillar structures. In addition, three-dimensional Floquet unit-cell dispersion analysis is conducted to investigate the underlying resonant behaviour of the periodic lattice. The results reveal nearly flat resonant branches and defect-sensitive frequency-selective transmission behavior consistent with locally resonant elastic metamaterial characteristics.

## 2. Numerical Modeling and Geometry

The geometry of the proposed periodic micro-pillar lattice is shown in **Figure 1.** The elastic-wave behavior of the structure was investigated using the Finite Element Method (FEM) implemented in COMSOL Multiphysics.

$$\nabla \cdot \sigma + f = \rho \frac{\partial^2 u}{\partial t^2} \tag{1}$$

where ρ is the material density, **u** is the displacement vector field, and **σ** represents the Cauchy stress tensor.

$$\sigma = C\varepsilon \tag{2}$$

Here, C represents the elastic stiffness tensor and ε the strain tensor.

$$\varepsilon = \frac{1}{2}[\nabla u + (\nabla u)^T] \tag{3}$$

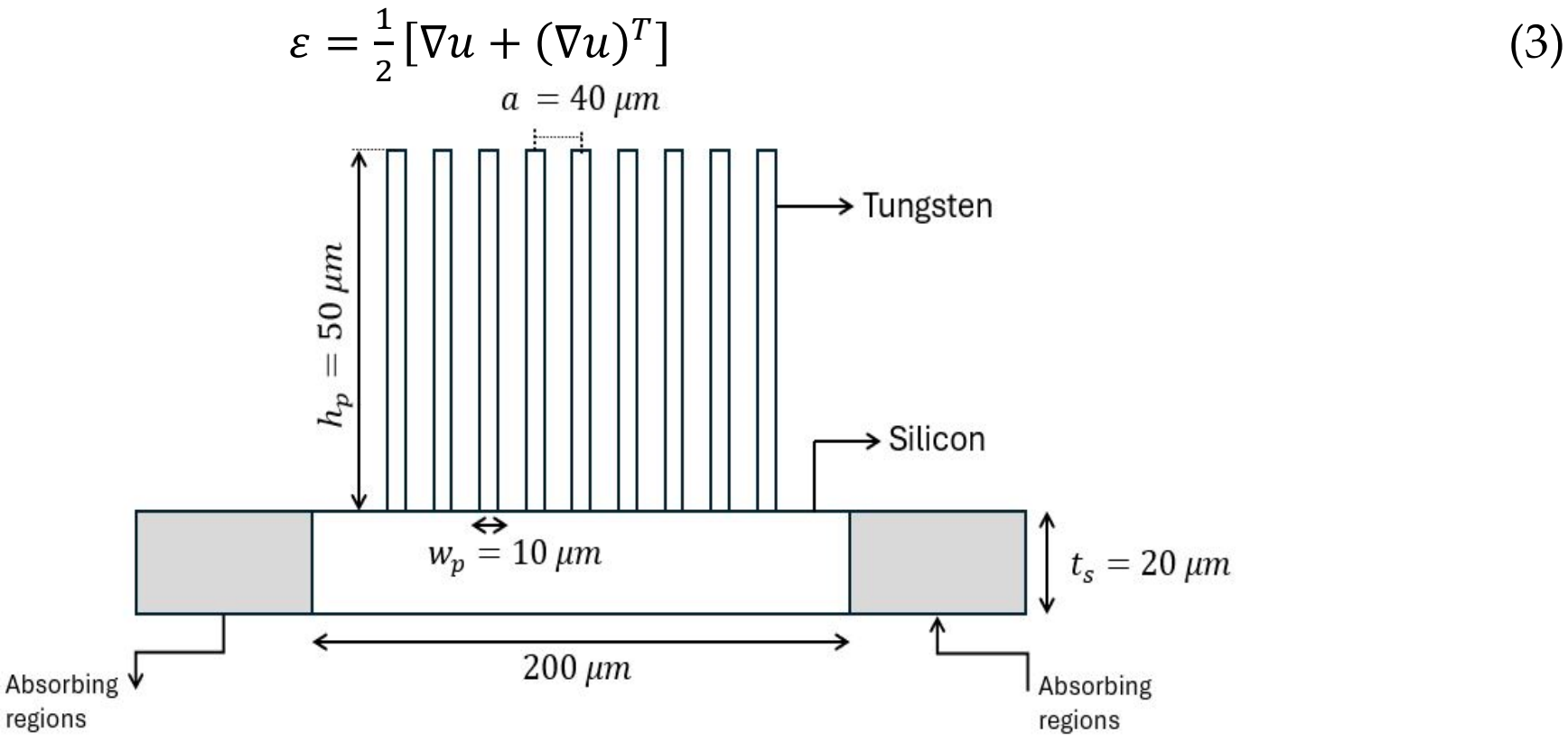

Figure 1. Schematic representation of the finite periodic micro-pillar lattice integrated on a silicon substrate used for the finite-domain transmission simulations. Tungsten resonators of height ($h_p$= 50μm) and width ($w_p$=10μm) are periodically distributed with lattice constant (a=40μm) on a silicon

slab of thickness (20μm). Absorbing regions are introduced near the domain boundaries to reduce artificial elastic-wave reflections during the frequency-domain simulations.

The strain tensor is obtained from the spatial displacement gradients under the assumption of small elastic deformations [9]. Both two-dimensional and three-dimensional models were considered in order to examine the resonant transmission behavior and the underlying dispersion characteristics of the periodic resonator lattice.

The structure consists of periodically distributed tungsten micro-pillars integrated on a silicon substrate. The lattice constant of the structure was chosen as ($a = 40\ \mu m$), while the square micropillars have the lateral dimensions of ($10 \times 10\ \mu m^2$) and a height of ($50\ \mu m$). The supporting silicon substrate thickness was fixed at ($20\ \mu m$).

$$\varphi = \frac{w_p^2}{a^2} \tag{4}$$

The corresponding geometrical filling factor of the periodic lattice is therefore given by ($\varphi = w_p^2/a^2 = 0.0625$), where $w_p$ denotes the pillar width.

For the two-dimensional studies, a side-view representation of the periodic structure was employed to efficiently investigate finite-domain resonant transmission behavior and defect-induced mode redistribution. The finite structures consisted of periodic arrays containing intentionally introduced missing-pillar defects with different defect widths corresponding to one, two, and three removed resonators. Frequency-domain simulations were carried out using harmonic out-of-plane prescribed displacement excitation applied at one side of the substrate, while absorbing damping regions were introduced near the domain boundaries in order to reduce artificial wave reflections.

$$u(r,t) = U(r)e^{i\omega t} \tag{5}$$

Here, $\omega$ denotes the angular excitation frequency and $U(r)$ represents the spatial displacement amplitude distribution. To validate the finite two-dimensional observations in realistic geometries, full three-dimensional finite array simulations were subsequently performed using a $9 \times 3$ periodic micro-pillar lattice integrated on a silicon slab. A central missing-pillar defect was introduced to investigate the influence of the defect-induced resonant transmission redistribution in three dimensions. The transmitted elastic wave response was quantified using surface-averaged displacement magnitudes evaluated at the output boundary of the structure.

In addition to the finite-array simulations, three-dimensional unit-cell Floquet dispersion analysis was conducted in order to investigate the underlying resonant behavior of the periodic lattice. Floquet periodic boundary conditions were applied along the in-plane directions of the unit-cell, while eigenfrequency analysis was performed to compute the dispersion characteristics along the high symmetry propagation direction of the Brillouin zone. The resulting dispersion curves revealed nearly flat resonant branches within the frequency range corresponding to the finite-array transmission modulation.

The silicon substrate and metallic resonators were modeled using isotropic linear elastic constitutive relations. Material parameters corresponding to silicon, tungsten, and copper were employed in the simulations in order to investigate the influence of resonator density and elastic contrast on the observed resonant transmission behavior.

## 3. Dispersion Analysis

The dispersion characteristics of the periodic micro-pillar lattice were investigated using three-dimensional Floquet unit-cell analysis in order to examine the resonant elastic wave behavior of the structure.

$$u(r+a) = u(r)e^{ik\cdot a} \tag{6}$$

Here, $k$ represents the Bloch wave vector and $a$ denotes the lattice periodicity vector of the unit cell [5]. A single periodic unit cell consisting of a tungsten micro-pillar integrated on a silicon substrate was considered as shown in **Figure 2,** and Floquet periodic boundary conditions were applied along the in-plane directions of the lattice. Eigen frequency analysis was subsequently performed along with the high symmetry propagation direction of the irreducible Brillouin zone [5–6].

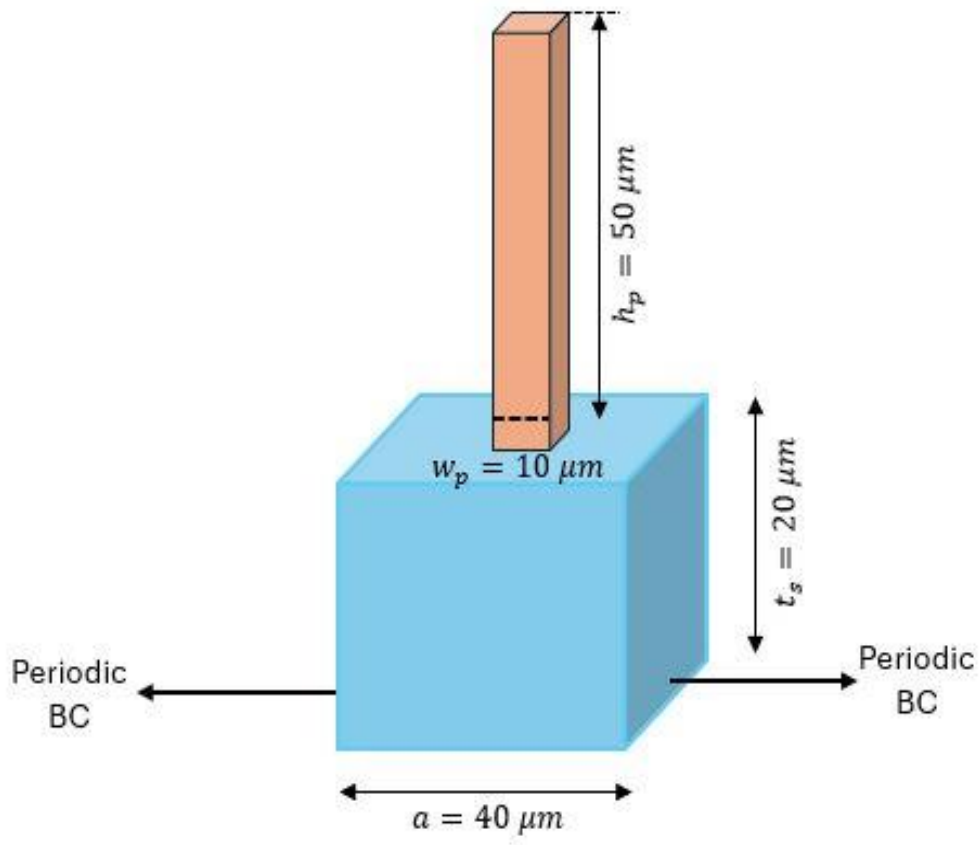

Figure 2. Three-dimensional Floquet unit-cell geometry of the periodic tungsten micro-pillar lattice integrated on a silicon substrate. Floquet periodic boundary conditions were applied along the in-plane directions of the unit cell in order to compute the elastic-wave dispersion characteristics of the periodic structure.

The calculated dispersion curve as presented in **Figure 3,** revealed the presence of several nearly dispersionless branches within the frequency range corresponding to the finite-array transmission studies.

$$v_g = \frac{d\omega}{dk} \tag{7}$$

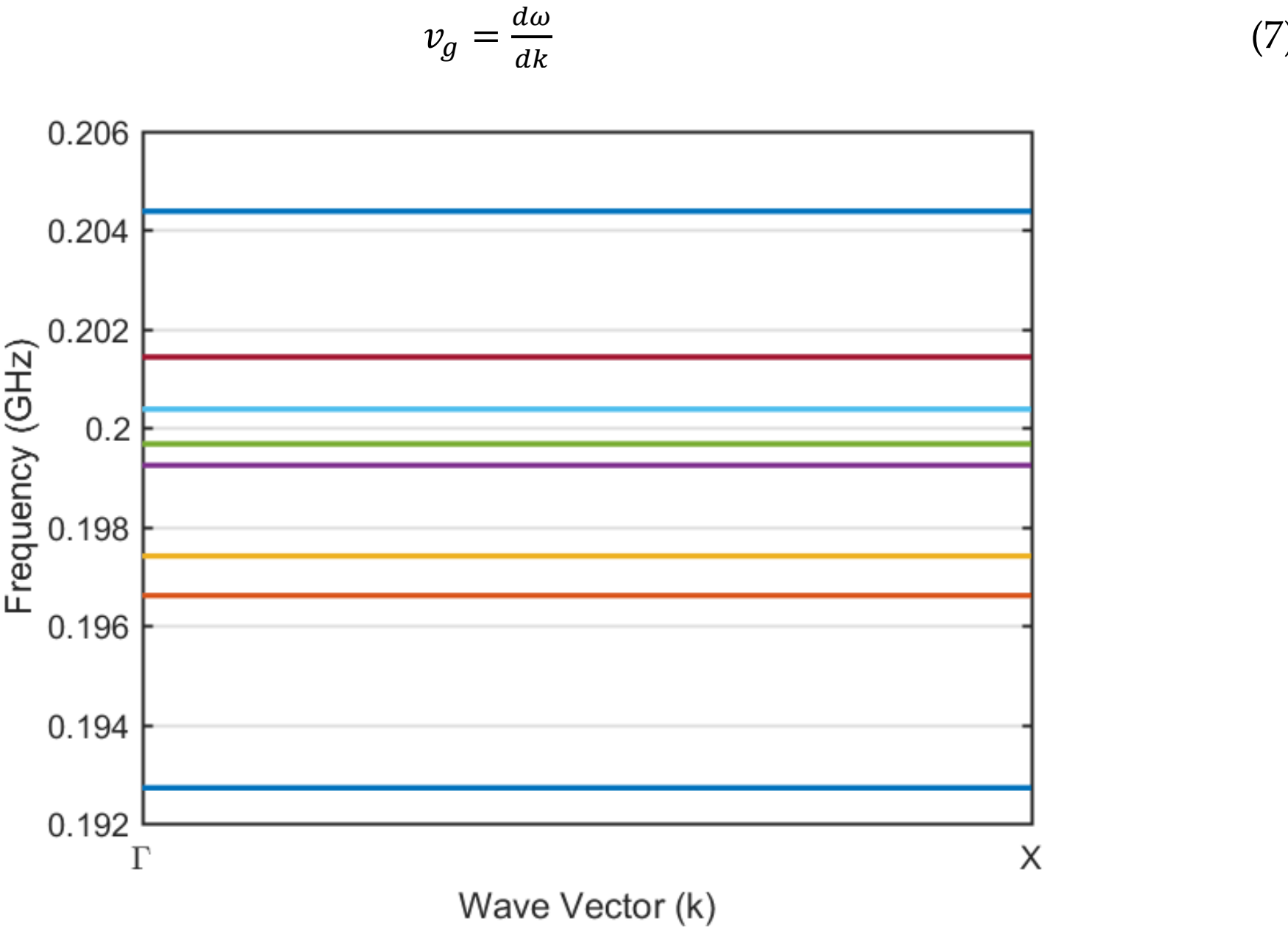

Figure 3. Floquet dispersion spectrum of the periodic tungsten micro-pillar unit cell integrated on a silicon substrate. Multiple nearly flat resonant branches are observed within the investigated frequency range, indicating low-group-velocity elastic-wave behavior associated with localized resonant interactions between the micro-pillars and the supporting substrate.

Nearly flat dispersion branches correspond to low group velocities, indicating weakly propagating locally resonant elastic-wave modes within the periodic micro-pillar lattice. Such flat resonant branches indicate low-group velocity elastic-wave behavior associated with the localized resonant interactions between the micro-pillars and the supporting substrate [3]. In contrast to strongly dispersive propagating branches typically observed in conventional Bragg-type phononic crystals, the present structure exhibits dominant locally resonant characteristics resulting from the periodic resonator array.

In addition to the flat resonant branches, sparse-mode frequency regions were observed within the dispersion spectrum, indicating reduced propagating elastic-wave states in specific frequency intervals. These stopband-like regions occur in the vicinity of the frequency range where strong transmission modulation was observed in the finite-array simulations. The correspondence between the nearly flat resonant bands and the defect-sensitive transmission behavior suggests that the observed elastic-wave transport is primarily governed by localized resonant interactions rather than purely Bragg-scattering induced bandgap formation.

The calculated dispersion behavior is therefore consistent with the interpretation of the proposed periodic micro-pillar structure as a locally resonant elastic metamaterial supporting frequency-selective elastic-wave transmission and resonant mode redistribution in the presence of structural defects.

## 4. Defect-Induced Resonant Transmission

The influence of structural defects on the elastic-wave transmission characteristics of the periodic micro-pillar lattice was investigated using finite-domain frequency-response simulations.

$$T_{norm}(f) = \frac{T(f)}{max[T(f)]} \tag{8}$$

The transmission response was normalized with respect to the maximum transmitted displacement amplitude within the investigated frequency range. Defect engineering was introduced by intentionally removing one, two, and three neighboring micro-pillars from the periodic array in order to examine the effect of defect width on resonant elastic-wave transport [4].

The frequency-domain transmission response of the perfect lattice exhibited multiple resonant transmission peaks within the investigated frequency range. Upon introducing missing-pillar defects, significant modifications of the transmission spectrum were observed, including the shifts in resonant frequencies, redistribution of transmission amplitudes, and the emergence of localized resonant transmission channels. These results indicate that the defect geometry strongly affects the resonant coupling between the periodic resonators and the elastic waves propagating through the supporting substrate.

To further visualize the defect-dependent transmission behavior, a transmission mapping analysis, shown in **Figure 4,** was performed as a function of excitation frequency and defect width.

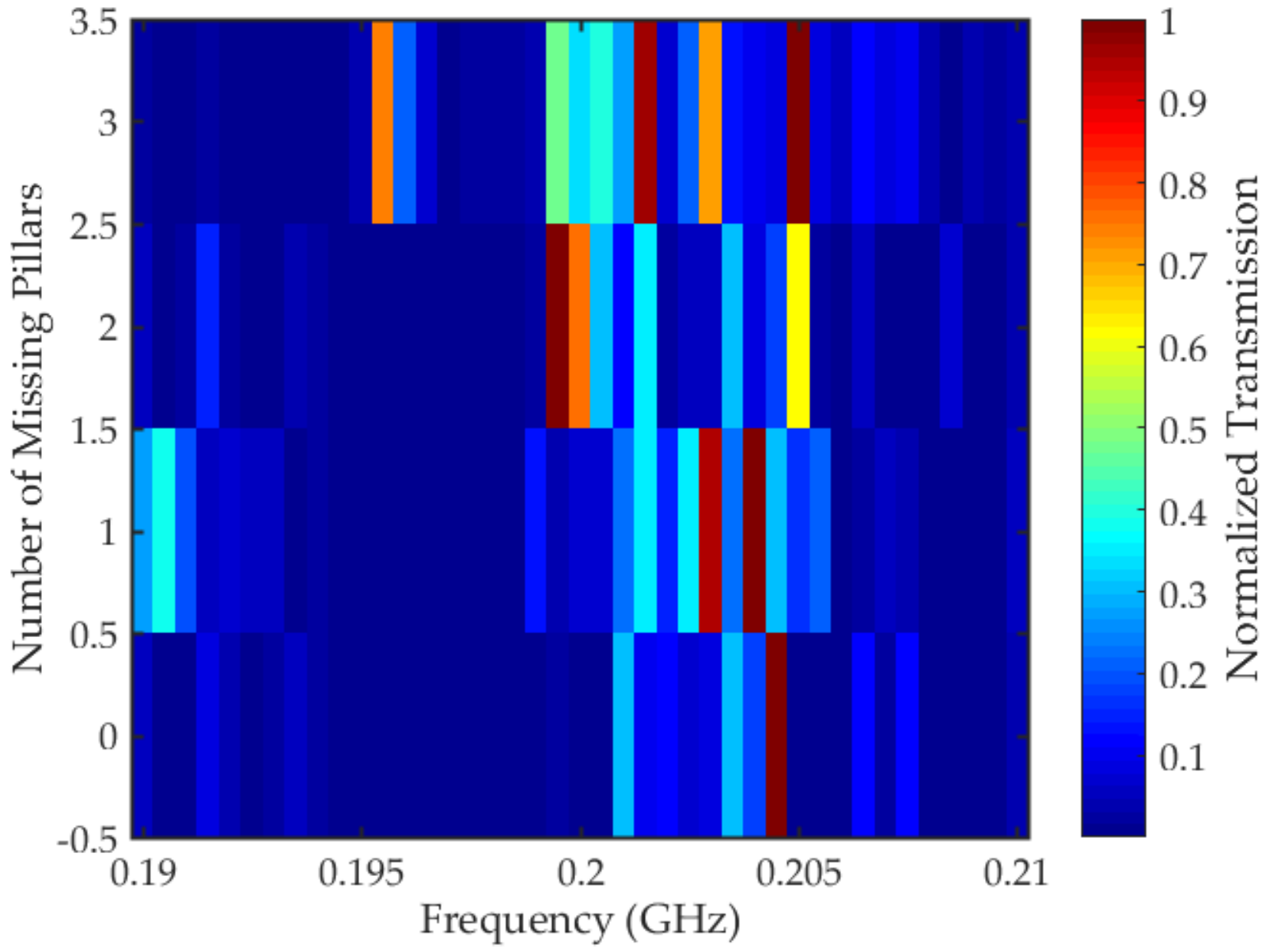

Figure 4. Transmission mapping analysis of the finite periodic micro-pillar lattice as a function of excitation frequency and defect width corresponding to different numbers of missing pillars. The results demonstrate strong defect-sensitive resonant elastic-wave transmission modulation together with frequency-selective redistribution of localized resonant states within the investigated frequency range.

The resulting transmission maps clearly demonstrated that increasing the number of missing pillars modifies both the location and intensity of the resonant transmission peaks. In particular, distinct frequency-selective transmission regions emerged for different defect configurations, revealing a strong dependence of the elastic-wave transport characteristics on the spatial extent of the defect region.

The displacement field distributions shown in **Figure 5,** obtained from the simulations further confirmed the presence of localized resonant elastic-wave behavior around the defect regions. Depending on the excitation frequency and defect width, strong displacement localization was observed near the missing-pillar cavities together with redistribution of vibrational energy among the neighboring resonators. These observations indicate that the introduced defects break the periodic symmetry of the lattice and modify the local resonant interactions governing elastic-wave propagation.

The observed transmission behavior therefore demonstrates that missing-pillar defects can effectively tune resonant elastic-wave transport in periodic micro-pillar lattices. Rather than exhibiting a rigorously proven complete phononic bandgap, the present structure primarily displays defect-sensitive locally resonant transmission modulation associated with low-group-velocity resonant branches identified in the dispersion analysis.

## 5. Three-Dimensional Validation

In order to verify the persistence of the observed resonant transmission behavior in realistic geometries, full three-dimensional finite-array simulations were subsequently performed using periodic micro-pillar lattices integrated on a silicon substrate. While the two-dimensional side-view models provide an efficient framework for investigating resonant transmission modulation and defect-induced localization phenomena, the three-dimensional simulations allow a more physically representative evaluation of the elastic-wave behavior within finite micro-pillar structures.

The three-dimensional lattice consisted of a $9 \times 3$ array of tungsten micro-pillars periodically distributed on a silicon slab. A central missing-pillar defect was introduced in order to investigate the influence of localized structural perturbations on the elastic-wave transmission response. Frequency-domain simulations were carried out within the same frequency range investigated in the two-dimensional studies, and the transmitted response was evaluated using surface-averaged displacement magnitudes at the output boundary of the structure.

The calculated three-dimensional responses revealed that the defect-induced transmission modulation observed in the two-dimensional simulations as shown in **Figure 5,** persists in the full three-dimensional geometry. In particular, the missing-pillar defect significantly modified the resonant transmission characteristics of the lattice, resulting in shifts of transmission peaks together with the redistribution of resonant elastic wave energy throughout the periodic structure. The observed transmission behavior confirms that the localized resonant interactions governing the finite-array transport are not restricted to simplified two-dimensional approximations and remain physically relevant in realistic three-dimensional micro-pillar configurations.

The displacement field distributions obtained from the three-dimensional simulations further demonstrated strong defect-sensitive resonant behavior within the periodic lattice.

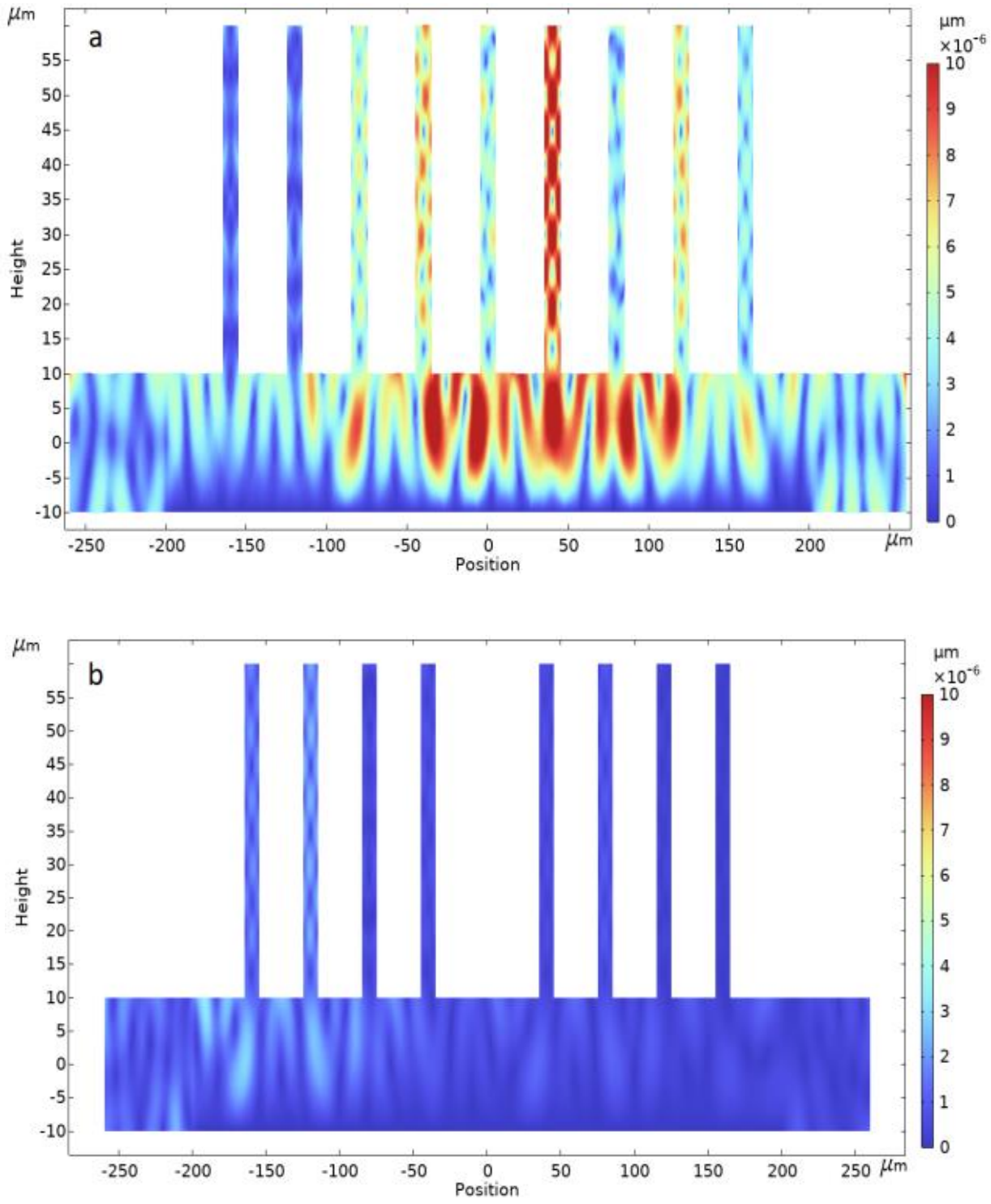

Figure 5. Displacement-field distributions of the finite periodic lattice at 0.201 GHz for (a) the perfect periodic array (top) and (b) the defect-containing array with missing pillars (bottom). The defect structure exhibits localized resonant elastic-wave redistribution and modified vibrational localization compared with the perfect lattice.

Depending on the excitation frequency, localized vibrational energy redistribution was observed near the defect region together with modified resonant excitation of neighboring pillars. Compared with the perfect lattice configuration, the defect structures exhibited altered spatial localization patterns and modified resonant transport pathways, indicating that the introduced defects significantly perturb the local resonant coupling mechanisms within the periodic array.

Overall, the three-dimensional simulations validate the robustness of the defect-induced resonant transmission behavior identified in the two-dimensional studies and confirm that the proposed periodic micro-pillar structures behave as locally resonant elastic metamaterials supporting defect-sensitive frequency-selective elastic-wave transport.

## 6. Material Comparison

The influence of resonator material properties on the elastic-wave transmission characteristics of the periodic micro-pillar lattice was further investigated through a comparative analysis between tungsten and copper resonators integrated on identical silicon substrates, as shown in **Figure 6**. The objective of this comparison was to evaluate the role of resonator density and elastic contrast on localized resonant behavior and defect-sensitive transmission modulation.

For both material configurations, the geometrical parameters and lattice arrangement were kept unchanged, while only the resonator material properties were modified. Frequency-domain simulations were subsequently performed for both the perfect and defect-containing finite lattices within the same excitation frequency range considered in the previous sections.

The calculated transmission responses revealed noticeable differences between the tungsten and copper resonator lattices. In particular, the tungsten-based structures exhibited more distinct and spatially localized resonant responses, indicating stronger local resonant interactions between the micro-pillars and the supporting substrate.

In contrast, the copper resonator structures displayed comparatively broader and more distributed vibrational responses throughout the lattice. Although defect-induced transmission modulation remained observable, the resulting resonant localization behavior was less pronounced than in the tungsten configuration. This behavior is attributed primarily to the lower density and reduced inertial contrast of copper relative to tungsten, resulting in weaker resonator-induced elastic-wave confinement.

The displacement field distributions further confirmed the stronger localized resonant behavior associated with the tungsten resonators.

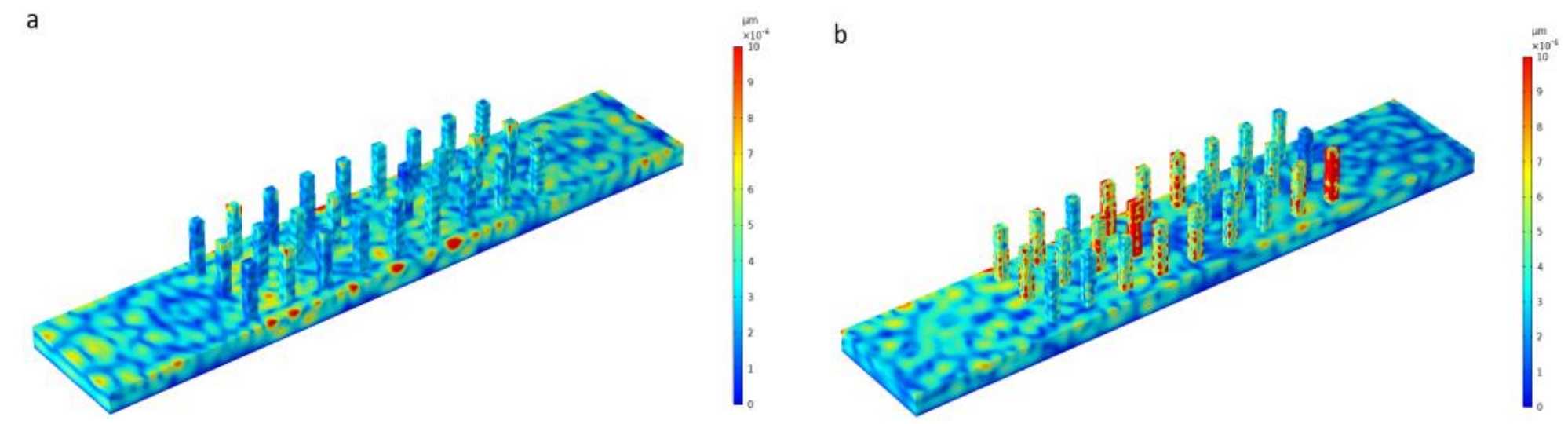

Figure 6. Three-dimensional displacement-field distributions of the defect-containing periodic micro-pillar lattice using (a) tungsten resonators (left) and (b) copper resonators (right) at 0.201 GHz. The tungsten-based structure exhibits stronger localized resonant elastic-wave behavior and enhanced vibrational confinement compared with the copper resonator lattice due to the larger inertial contrast between the resonators and the supporting silicon substrate.

In the tungsten structures, vibrational energy was more strongly concentrated near the resonant defect regions, whereas the copper lattices exhibited more spatially distributed elastic-wave propagation patterns. These observations indicate that increasing resonator density enhances localized resonant interactions and strengthens defect-sensitive transmission modulation within the periodic micro-pillar lattice.

Overall, the material comparison demonstrated that resonator density and elastic contrast play a significant role in governing the locally resonant elastic-wave transport characteristics of periodic micro-pillar structures. The stronger resonant localization observed in the tungsten-based lattices further supports the interpretation of the proposed structures as locally resonant elastic metamaterials dominated by resonator-induced transmission behavior.

## 7. Discussion

The present results demonstrate that periodic micro-pillar lattices integrated on thin elastic substrates can support strong defect-sensitive resonant elastic-wave transmission behavior governed primarily by localized resonant interactions. The combination of finite-array transmission simulations and Floquet dispersion analysis consistently indicates that the proposed structures behave predominantly as locally resonant elastic metamaterials rather than conventional Bragg-scattering-dominated phononic crystals.

The nearly flat dispersion branches observed in the unit cell Floquet analysis indicate low-group-velocity resonant behavior associated with weakly propagating elastic-wave modes. Such flat resonant bands are characteristic of locally resonant systems in which the resonator dynamics dominate the elastic-wave response of the structure. In the present micro-pillar lattice, the periodic resonators strongly influence the transmission characteristics by introducing localized vibrational states and resonant redistribution of elastic-wave energy throughout the substrate. The observed transmission behavior is interpreted predominantly in terms of locally resonant elastic wave interactions rather than classical Bragg-scattering-induced bandgap formation. Furthermore, the investigated range corresponds to a subwavelength regime relative to the lattice periodicity ($a=40\mu m$), supporting the interpretation that the observed transmission features arise predominantly from resonator-induced elastic-wave modulation rather than classical first-order Bragg-scattering bandgaps. Considering the relatively small lattice periodicity and the corresponding elastic wavelength within the investigated frequency range, the structure operates primarily within a subwavelength locally resonant regime characterized by nearly flat resonant branches and defect-sensitive transmission modulation.

The finite-domain simulations further revealed that introducing missing-pillar defects significantly modifies the resonant transmission characteristics of the lattice. Depending on the defect width, the resonant transmission peaks shift and redistribute within the investigated frequency range, indicating that the local resonant coupling mechanisms are highly sensitive to structural perturbations. The resulting defect-dependent transmission modulation demonstrates that the introduced cavities act as localized resonant perturbations capable of modifying the elastic-wave transport pathways within the periodic lattice.

Unlike conventional phononic crystal systems designed primarily for the formation of complete Bragg bandgaps, the present structure exhibits dominant localized resonant behavior together with stopband-like sparse-mode frequency regions rather than rigorously demonstrated complete phononic bandgaps. This behavior is consistent with the relatively low filling factor and moderate pillar-to-substrate height ratio employed in the proposed geometry, which favors localized resonant interactions over strong propagating-mode hybridization.

The material comparison additionally demonstrated that the resonator density plays an important role in determining the strength of the localized resonant behavior. The tungsten resonator lattices exhibited stronger resonant localization and more pronounced defect-sensitive transmission modulation than the copper structures due to larger inertial contrast introduced by the higher-density resonators. These observations further support the interpretation of the transmission behavior as being governed primarily by resonator-induced elastic-wave interactions.

## 8. Conclusion

In this work, defect-width-dependent resonant elastic-wave transmission in periodic micro-pillar lattices integrated on silicon substrates was numerically investigated using finite-element simulations. Two-dimensional finite-domain studies revealed that the introduction of missing-pillar defects significantly modifies the resonant transmission behavior of the periodic lattice, resulting in frequency-selective transmission redistribution and localized resonant elastic-wave interactions.

Three-dimensional finite-array simulations further confirmed the persistence of the observed defect-sensitive transmission behavior in realistic micro-pillar geometries. The calculated displacement field distributions demonstrated strong localized resonant behavior near the defect regions together with modified elastic-wave transport pathways throughout the periodic structure.

In addition, three-dimensional Floquet unit-cell dispersion analysis revealed the presence of nearly flat resonant branches and low-group-velocity elastic-wave behavior within the investigated frequency range. The correspondence between the dispersion characteristics and the finite-array transmission response indicates that the proposed structures behave predominantly as locally resonant elastic metamaterials rather than conventional Bragg-scattering-dominated phononic crystals.

The comparative analysis between tungsten and copper resonator lattices further demonstrated the important role of resonator density and elastic contrast in governing the strength of the localized resonant behavior and defect-sensitive transmission modulation. The tungsten resonators exhibited stronger resonant localization and more pronounced transmission redistribution due to their larger inertial contrast relative to the silicon substrate.

Overall, the present study demonstrates that missing-pillar defects provide an effective mechanism for tuning resonant elastic-wave transport in periodic micro-pillar lattices. The observed defect-width-dependent transmission behavior may provide useful perspectives for the development of frequency-selective elastic-wave control, localized vibration management, and defect-engineered elastic metamaterial structures operating at microscale dimensions.